\documentclass[12pt]{article}
\usepackage{setspace}
\usepackage{color}
\usepackage{graphics}

\begin{document}

\newcommand{\ra}{\rangle }
\newcommand{\la}{\langle }
\newcommand{\ket}[1]{| #1 \rangle }
\newcommand{\bra}[1]{\langle #1 | }
\newcommand{ \ave}[2]{ \langle #1 | #2 | #1 \rangle }
\newcommand{\amp }[2]{\langle #1|#2 \rangle }
\newcommand{\weakv}[3]{ \frac{\langle #1|#2| #3\rangle}{\langle #1 | #3 \rangle }}
\newcommand{\beq}{\begin{equation}}
\newcommand{\eeq}{\end{equation}}
\newcommand{\up}[1]{^{(#1)}}
\newcommand{\upn}{^{(N)}}
\newcommand{\upa}{\uparrow}
\newcommand{\dwa}{\downarrow}
\newcommand{\beqa}{\begin{eqnarray}}
\newcommand{\eeqa}{\end{eqnarray}}
\newcommand{\beqar}{\begin{eqnarray*}}
\newcommand{\eeqar}{\end{eqnarray*}}
\newcommand{\be}{\beta}
\newcommand{\al}{\alpha}
\newcommand{\ie}{{\it i.e.,}\ }
\newcommand{\eg}{{\it e.g.,}\ }
\newcommand{\ssc}{\scriptscriptstyle}
\def \vr {\hat\varrho}
\def \vra {\varrho}
\def \bvra {{\bf\varrho}}
\def \prob {{\cal P}rob}
\def \la {\langle}
\def \ra {\rangle}
\def \up {\uparrow}
\def \down {\downarrow}
\def \o {{\cal O}}
\def \h {{\cal H}}
\def \s {{\cal S}}
\def \r {{\cal R}}
\def \e {{\cal E}}
\def \md {{\cal MD}}
\def \a {\alpha}
\def \b {\beta}
\def \c {\gamma}
\def \d {\delta}
\def \hl {\Delta_L}
\def \hr {\Delta_R}
\def \u {\uparrow}
\def \d {\downarrow}
\def \P {\cal P}

\newcommand{\eps} {\varepsilon}
\newcommand{\sig} {\hat{\sigma}}
\newcommand{\rhod} {\rho\mathrm{^{DEN}}}
\newcommand{\half} {\frac{1}{2}}
\newcommand{\hrt} {\frac{1}{\sqrt{2}}}
\newcommand{\ox} {\otimes}
\newcommand{\imp} {\longrightarrow}
\newcommand{\uk} {\ket{\up}}
\newcommand{\scal} [2] {\langle#1\vert#2\rangle}
\newcommand{\dk} {\ket{\dn}}
\newcommand{\ub} {\bra{\up}}
\newcommand{\db} {\bra{\dn}}
\newcommand{\ak} [1] {\ket{\mathrm{#1}}}
\newcommand{\ab} [1] {\bra{\mathrm{#1}}}
\newcommand{\sog} [1] {\left(#1\right)}
\newcommand{\ek} [2] {\ket{\mathrm{\eps_{#1}\sog{#2}}}}
\newcommand{\eb} [2] {\bra{\mathrm{\eps_{#1}\sog{#2}}}}
\newcommand{\abs} [1] {\vert#1\vert}
\newcommand{\si} [1] {\sin\sog{#1}}
\newcommand{\co} [1] {\cos\sog{#1}}
\newcommand{\dn} {\downarrow}
\renewcommand{\theequation}{\arabic{section}.\arabic{equation}}

\centerline{\large \bf Non-statistical Weak Measurements}

\bigskip

\centerline{\bf Jeff\footnote{Corresponding Author. Address: 4400 University Drive MSN 5C3, Fairfax, VA 22030. Phone: 703-993-4322, email: jtollaks\@gmu.edu} Tollaksen$^{(a)}$, Yakir Aharonov$^{(a), (b), (c)}$}

\bigskip

\centerline{(a) Department of Physics and Department of Computational and Data Sciences, }
\centerline{College of Science, George Mason University, Fairfax, VA 22030}
\centerline{(b) School of Physics and Astronomy, Tel
 Aviv University, Tel Aviv 69978, Israel} 
\centerline{(c) Department of Physics, University of South Carolina, Columbia, SC 29208.}
\begin{abstract}
Non-statistical weak measurements  yield  weak values that are outside the range of eigenvalues and are not rare, suggesting that weak values are a property of every pre-and-post-selected ensemble.  They also extend the applicability and valid regime of weak values.
\end{abstract}


\vskip -1cm

%\newpage
\section{\textcolor{black}{\bf  Introduction}}

%\vskip -1cm
Aharonov, Bergmann and Lebowitz
(ABL,~\cite{abl}) considered  measurement situations {\em between} two successive ideal measurements where the transition from a pre-selected state $\ket{\Psi_{\mathrm{in}}}$ to a post-selected state $\ket{\Psi_{\mathrm{fin}}}$ is  generally disturbed by an intermediate precise measurement.  
A subsequent theoretical development arising out of the ABL work was the introduction of the ``Weak Value" (WV) of an observable which was probed by a  new type of quantum measurement called the ``Weak Measurement" (WM) \cite{av} (reviewed in \S \ref{der2v}). 
The motivation behind these measurements was to explore the relationship between $\ket{\Psi_{\mathrm{in}}}$ and $\ket{\Psi_{\mathrm{fin}}}$  by reducing the disturbance on the system during the intermediate time.  For example, if a WM of $\hat{A}$ is performed at the intermediate time $t$ ($t_{\mathrm{in}}<t<t_{\mathrm{fin}}$) then, in contrast to the ABL situation,  the basic object in the entire interval $t_{\mathrm{in}}\rightarrow t_{\mathrm{fin}}$ for the purpose of calculating {\it other} WVs for other measurements 
is the pair of states $\ket{\Psi_{\mathrm{in}}}$ and $\ket{\Psi_{\mathrm{fin}}}$.   However, the reduction of disturbance also reduced the information obtained from a single WM on a single  quantum system.  Therefore, it was believed that the WV could only be determined ``precisely" by using a statistical approach and a large ensemble (reviewed in \S \ref{wvstat} and \S\ref{wvavgop}).
This was a result of the weakness condition which produced a shift in the pointer of the measuring device (MD) that was much less than it's uncertainty.
Many separate irreversible recordings of the slight MD shift were then used to amplify the ``weak value signal" above the ``noise" due to the weakened measurement.

This letter introduces (\S \ref{rwm}) a new Gedanken experiment coined ``non-statistical weak measurement" (NSWM)
which involves an irreversible recording of  the sum of momenta (i.e. it's center of mass), such that it's shift (which precisely registers a WV) is large compared to it's noise. Thus, with NSWM, the ``weak value signal" is greater than the noise in just a single weak measurement, unlike the former approach.  
In addition, in order to ascertain the maximally allowed information about the WV, the relative positions (which commute with the total momenta) are also measured.  
In a physical, ``realistic," WM,  there is always a finite coupling and thus a disturbance caused to the system.
However, we can use the relative positions to correct for this disturbance and thus WVs can now be determined accurately even with a finite-sized ensemble.  Therefore, 
by increasing the interaction strength between MD and system to a comparatively large value, we  expand the validity of WVs outside the domain of it's original context~\cite{at}.  
NSWM is not statistical because we are making only one single measurement rather than many.  
In a subsequent publication, we will consider the possibility of a realistic experimental realization of NSWM~\cite{abt} which holds the practical application of amplifying weak or unknown signals.
Finally,
we also demonstrates how eccentric weak values (EWV), i.e. WVs outside the eigenvalue spectra, can be obtained in an ensemble which is not rare.  
We shall focus on EWVs because such results cannot be reproduced from a positive definite probability distribution of eigenvalues and therefore hold the possibility of revealing new aspects of Quantum Mechanics (QM).

\label{tpwm}

\vskip -.5cm
\section{Weak Measurements}

\label{der2v}

It will be useful to briefly review the derivation of WVs and WMs~\cite{av,duck}.
WMs can be quantified in the quantum measurement theory developed by von Neumann~\cite{vn}.
First we consider an ideal measurement of observable $\hat{A}$ by using an interaction Hamiltonian $H_{\mathrm{int}}$ of the form 
$H_{\mathrm{int}}=-\lambda  (t)\hat{Q}\hat{A}$
where $\hat{Q}$ is an observable of the MD and $\lambda (t)$ is a coupling constant which is non-zero only during a short time $(0,T)$.  
Using the Heisenberg equations of motion for the momentum $\hat{P}$ of MD (conjugate to the position $\hat{Q}$), we see that $\hat{P}$ changes according to $\frac{\d\hat{P}}{\d t}=\lambda  (t) \hat{A}$.
Integrating this, we see that $P(T)-P(0)=\lambda A$
where $\int_0^T \lambda  (t)dt=\lambda $.  
To make a more precise determination of $\hat{A}$ requires that $P(0)$ and $P(T)$ are more precisely defined, which occurs, e.g. if MD approaches a delta function in $P$.
However, in this case the disturbance or back-reaction on the system is increased due to a larger  $H_{\mathrm{int}}$ which is a result of the larger $\Delta Q$.
When $\hat{A}$ is 
measured then any $\hat{B}$ (not commuting with $\hat{A}$) is disturbed because $\frac{d}{dt}{\hat{B}}=\frac{{\mathrm{i}}}{\hbar}\lambda  (t)[\hat{A},\hat{B}]\hat{Q}$, and since $[\hat{A},\hat{B}]\neq 0$ and the spread of $\hat{Q}$ is not zero, $B$ changes in an uncertain way proportional to $\lambda\Delta Q$.  

For WMs, we can use von Neumann, except that we post-select the system and weaken the interaction using one of the following criteria:  
\begin{enumerate}
\item use a small $\lambda$ (assuming without lack of generality that the state of the MD is a Gaussian with spreads $\Delta\equiv\Delta P=\Delta Q=1$), or 

\item  $\hat{P}$ of MD is measured to a finite
 precision
 $\Delta P$, (which limits the disturbance  by a
 finite amount $\Delta Q\geq 1/\Delta P$), or
\item  $[A^{n}_{\mathrm{w}} -
(A_{\mathrm{w}})^{n}]/n!$ is small ($(A^n)_{\mathrm{w}} \equiv { {\langle \Psi _{\mathrm{fin}} \!\mid A^n \mid \Psi _{\mathrm{in}}
\rangle} \over {\langle \Psi _{\mathrm{fin}} \!\mid \Psi _{\mathrm{in}}
\rangle}}$), even if $\lambda\hat{Q}$ is not small.   
\end{enumerate}
The simplest derivation of the WV result is with the first approach, i.e. $\lambda$ small ($\int \lambda  (t)dt=\lambda \ll 1$), in which there is little disturbance or back-reaction on the system.  The final state of MD after WM and post-selection is: 
\begin{eqnarray}
 \tilde{\Phi}_{\mathrm{fin}}^{\mathrm{MD}}(P) & \to &  \bra{\Psi_{\mathrm{fin}}}e^{ -i \lambda  \hat{Q} \hat{A}
 }\ket{\Psi_{\mathrm{in}}}\tilde{\Phi}_{\mathrm{in}}^{\mathrm{MD}}(P)\approx\langle\Psi_{\mathrm{fin}}\!\mid
\Psi_{\mathrm{in}} \rangle \lbrace 1+i\lambda \hat{Q} A_{\mathrm{w}})\tilde{\Phi}_{\mathrm{in}}^{\mathrm{MD}}(P)\nonumber\\
&\approx &\langle\Psi_{\mathrm{fin}}\ket{\Psi_{\mathrm{in}}}e^{ -i \lambda  \hat{Q}A_{\mathrm{w}}
 }\tilde{\Phi}_{\mathrm{in}}^{\mathrm{MD}}(P)\approx\exp\left\{{-{{(P-\lambda \,
 A_{\mathrm{w}})^2}
}}\right\}\\
where \,\,\,A_w&=&\weakv {\Psi_\mathrm{fin}}{\hat{A}}{\Psi_\mathrm{in} }
\label{wv1}
 \label{post_selected}
\end{eqnarray}
The final
 state
 of MD is almost un entangled with the
 system and is shifted by a very surprising amount, the WV, $A_{\mathrm{w}}$.
We have used such limited
 disturbance measurements to explore many 
 paradoxes (see, e.g. \cite{at2}).  There have also been a number  of experiments to test the predictions made by the WM and their results are in very good agreement with
 theoretical predictions \cite{RSH,Ahnert,Pryde,Wiseman,Parks}.  

Because the new NSWM introduced in \S \ref{rwm} 
uses components of each of these 3  criterion, it will also be useful to review criterion 2 and 3 in which  $\hat{P}$ of MD is measured to a finite
 precision
 $\Delta P$.
In this regime, the measurement becomes
 less
 precise because the uncertainty $\Delta P$ in the position of
 the pointer is larger than the difference in the shifts of
 the pointer $\lambda a_i$ corresponding to the different
 eigenvalues and thus the shift in MD is much smaller than its uncertainty.
The final
 state
 of MD is almost un-entangled with the
 system and is centered on the 
 value $\bar A=\la\Psi|A|\Psi\ra$.  This can be seen by considering the measurement process in the Schroedinger picture:
\begin{equation}
|\Phi_{tot}\ra = |\Psi_{in}\ra |\Phi_{in}^{MD}\ra\rightarrow e^{ -i  \int H_{int} dt }|\Psi_{in}\ra |\Phi_{in}^{MD}\ra=e^{ i  \lambda\hat{Q} \hat{A}}|\Psi_{in}\ra |\Phi_{in}^{MD}\ra
\label{measurement2}
\end{equation}
where  the state of the system is $|\Psi_{in}\ra$ and the MD state, $|\Phi_{in}^{MD}\ra$, is given by $|\Phi_{in}^{MD}\ra=\int dQ \Phi_{in}^{MD}(Q)|Q\ra=\int dP \tilde{\Phi}_{in}^{MD}(P)|P\ra$.
A good approximation for realistic experiments is  to consider MD's initial state as a Gaussian (without loss of generality), i.e. 
 $\Phi_{in}^{MD}(Q)\equiv\la Q|\Phi_{in}^{MD}\ra = \exp(-{{Q^2}\over{4\Delta^2}})$ and $\tilde{\Phi}_{in}^{MD}(P)\equiv\la P|\Phi_{in}^{MD}\ra = \exp(-\Delta^2P^2)$ (substituting $\Delta\equiv\Delta
Q $,  $\Delta P\equiv\frac{1}{2\Delta}$, leaving off the normalizations and setting $\hbar=1$).  Expanding $|\Psi_{in}\ra$ in eigenstates of $\hat{A}$, i.e. $|\Psi_{in}\ra =\sum_i|A=a_i\ra\la
 A=a_i|\Psi_{in}\ra =\sum_i \alpha_i|A=a_i\ra$
then eq. \ref{measurement2}, becomes:
\begin{equation}
  \sum_i\alpha_i\int dQ e^{i\lambda Qa_i}e^{-{{Q^2}\over{4\Delta^2}}}|A=a_i\ra|Q\ra =\sum_i\alpha_i\int dP e^{{-(P-\lambda a_i)^2}\Delta^2}|A=a_i\ra|P\ra 
 \label{measurement3}
\end{equation}
When the uncertainty $\Delta P=\frac{1}{2\Delta Q}$ in the pointer is
 much smaller than the difference in the final shifts of the pointer 
 corresponding to different eigenvalues $a_i$, then eq. \ref{measurement3} has all the characteristics of an ideal measurement: the final state of the
 pointer
 (after tracing over the state of the measured system) is a
 density
 matrix representing a series of peaks, each corresponding to
 a
 different eigenvalue $a_i$ (because ${ \Phi_{fin}^{MD}(P-\lambda a_i)}\bot { \Phi_{fin}^{MD}(P-\lambda a_j)}$ when
$i\neq j$), and having probability equal to
 $|\la
 A=a_i|\Psi_{in}\ra|^2$.

However, in the weak regime ($\Delta P\gg \lambda a_i$), the final state of MD is a superposition of many substantially overlapping Gaussians with a distribution $Prob(P)=\sum_i |\la A=a_i|\Psi_{in}\ra|^2 e^{-\frac{(P-\lambda a_i)^{2}} {2\Delta P^{2}}} $, which ends up being a single Gaussian 
 $\Phi_{\mathrm{fin}}^{\mathrm{MD}}\approx\exp(-{{(P-\bar
 A)^2}\over{\Delta P^2}})$ centered on $\bar{A}$.
Traditionally, this average $\bar{A}$ is obtained statistically because each measurement yields an eigenvalue and thus we need to repeat the measurement many times to
 be able to
 locate the center. 

If we now add a post-selection to this ordinary, weakened, von Neumann measurement, then unique aspects of QM, namely the WV, can be probed.
As an indication of this, consider inserting a complete set of states $\{ \ket{\Psi_{\mathrm{fin}}}_j \}$ into 
$\bar{A}$:
\beq
 \bar{A} =  \bra{\Psi_{\mathrm{in}}} { \left[\sum_j  \ket{\Psi_{\mathrm{fin}}}_j\bra{\Psi_{\mathrm{fin}}}_j\right]\hat{A}} \ket{\Psi_{\mathrm{in}}}
= \sum_j |\langle \Psi _{\mathrm{fin}} \!\mid_j \!\Psi _{\mathrm{in}}
\rangle|^2\ 
{ {\langle \Psi _{\mathrm{fin}}\! \mid_j \hat{A} \mid \!\Psi _{\mathrm{in}}
\rangle} \over {\langle \Psi _{\mathrm{fin}} \!\mid_j \!\Psi _{\mathrm{in}}
\rangle}}
\label{expweak}
\eeq
If we interpret the states $ \ket{\Psi_{\mathrm{fin}}}_j $ as the possible outcomes of some  final measurement on the system (i.e. a post-selection at $t_{\mathrm{fin}}$, where for simplicity, we neglect the free time evolution of $ \ket{\Psi_{\mathrm{fin}}}_j $ and $\bra{\Psi_{\mathrm{in}}}$) and we perform an ideal WM (i.e. with $\lambda\Delta Q\rightarrow 0$) during the intermediate time $t<t_{\mathrm{fin}}$, then the coefficients $|\langle \Psi_{\mathrm{fin}}\! \mid_j \Psi_{\mathrm{in}}
\rangle |^2$ give  the probabilities $P(j)$ for obtaining a pre-selection of $\bra{\Psi_{\mathrm{in}}}$ and a post-selection of  $ \ket{\Psi_{\mathrm{fin}}}_j $ (since the intermediate WM does not disturb these states). $\bar{A}$ can thus be viewed as 
$\bar{A} = \sum_j P(j)\,  A_{\mathrm{w}}(j)$,
where the  quantity
$A_{\mathrm{w}}(j) \equiv { {\langle \Psi _{\mathrm{fin}} \!\mid_j \hat{A} \mid \!\Psi _{\mathrm{in}}
\rangle} \over {\langle \Psi _{\mathrm{fin}} \!\mid_j \!\Psi _{\mathrm{in}}
\rangle}}$
 is the WV of $\hat{A}$ given a particular final post-selection $\langle \Psi _{\mathrm{fin}} \!\mid_j$. 
Thus, one can think of  $\bar{A}$ for the whole ensemble as being built out of pre- and post-selected
states in which   the WV is multiplied by a probability for post-selection.

More precisely, if we expand the post-selected state $|\Psi_{\mathrm{fin}}\ra$ in eigenstates of $\hat{A}$, i.e. $
|\Psi_{\mathrm{fin}}\ra =\sum_i|A=a_i\ra\la
 A=a_i|\Psi_{\mathrm{fin}}\ra =\sum_i \beta_i|A=a_i\ra$ and (without loss of generality) choose a Gaussian state for MD, then the final state of MD in the position representation is (substituting $\Delta\equiv\Delta
Q $,  $\Delta P\equiv\frac{1}{2\Delta}$ and defining $(A^n)_{\mathrm{w}} \equiv { {\langle \Psi _{\mathrm{fin}} \!\mid A^n \mid \Psi _{\mathrm{in}}
\rangle} \over {\langle \Psi _{\mathrm{fin}} \!\mid \Psi _{\mathrm{in}}
\rangle}}$):

\begin{eqnarray}
\tilde{\Phi}_{\mathrm{fin}}^{\mathrm{MD}}(Q) &= & \sum_{n=0}^{\infty} \frac{-i(\lambda\hat{Q})^n}{n!} \langle\Psi_{\mathrm{fin}}\!\mid \hat{A}^n
\mid\Psi_{\mathrm{in}} \rangle e^{-{\hat{Q}^{2} \over {4\Delta^{2}}}}  
=\langle\Psi_{\mathrm{fin}}\!\mid \Psi_{\mathrm{in}}\ra\sum_{n=0}^{\infty} \frac{(-i\lambda\hat{Q})^n}{n!} (A^n)_{\mathrm{w}}
e^{-{\hat{Q}^{2} \over {4\Delta^{2}}}} \nonumber \\
&=& \langle\Psi_{\mathrm{fin}}\!\mid
\Psi_{\mathrm{in}} \rangle \lbrace 1+i\lambda \hat{Q} A_{\mathrm{w}} + \sum_{n=2}^{\infty}
{(i\lambda \hat{Q})^{n} \over n!} A^{n}_{\mathrm{w}} \rbrace e^{-{\hat{Q}^{2} \over {4\Delta^{2}}}} \nonumber\\
&=& \langle\Psi_{\mathrm{fin}}\!\mid \Psi_{\mathrm{in}} \rangle \lbrace 
1+i\lambda \hat{Q} A_{\mathrm{w}} +  \sum_{n=2}^{\infty} -
{(i\lambda \hat{Q})^{n} \over n!} (A_{\mathrm{w}})^{n} \nonumber\\
&\,\,\,\,\,\,\,\,& - \sum_{n=2}^{\infty}
{(i\lambda \hat{Q})^{n} \over n!} (A_{\mathrm{w}})^{n} +\sum_{n=2}^{\infty}
{(i\lambda \hat{Q})^{n} \over n!} (A^{n})_{\mathrm{w}} \rbrace e^{-{\hat{Q}^{2} \over {4\Delta^{2}}}} \nonumber \\
&=& \langle\Psi_{\mathrm{fin}}\!\mid \Psi_{\mathrm{in}}  \rangle \lbrace e^{\frac{-i}
{\hbar}\lambda \hat{Q}A_{\mathrm{w}}}  +  \sum_{n=2}^{\infty}
{(i\lambda \hat{Q})^{n} \over n!} [A^{n}_{\mathrm{w}} - (A_{\mathrm{w}})^{n}] \rbrace e^{-{\hat{Q}^{2} \over {4\Delta^{2}}}}
\label{mstate2}
\end{eqnarray}
The second term in the last part of eq. \ref{mstate2} (i.e. the
higher order 
terms) can be neglected if the second and/or third weakness criterion are met.  Then  eq. \ref{post_selected} (i.e. a shift of MD by the WV, $A_w$) is obtained when eq. \ref{mstate2} is transformed back to the $P$
representation.

Another result discussed in this letter is a new method to increase the coupling (i.e. $\lambda\sim 1$, not $\lambda \ll 1$ as used above).   
The difficulty with this is that the WV approximations become more and more precise in the idealized weak limit of $\lambda\Delta Q \rightarrow 0$ in which there is no disturbance or back-reaction on the system.  
We demonstrate a new way to increase $\lambda\Delta Q$ while maintaining the accuracy of the WV.  In order to accomplish this, we will consider the structure of MD in $Q$ (in addition to $P$) which registers information concerning the back reaction on the system.  
Intuitively, we can see 2 inverse roles for MD observable $\hat{Q}$ and system observable $\hat{A}$.  We have already reviewed how $\hat{Q}$ generates translations in $\hat{P}$ proportional to $\hat{A}$.  However, the roles for $\hat{Q}$ and $\hat{A}$ are reversed when one asks the reverse question: ``What is 
the back reaction on the {\it system} (i.e. not on the {\it MD}) due to the measurement interaction $H_{\mathrm{int}}$?"  For this question,  $\hat{A}$ is now the generator (not $\hat{Q}$) in a manner proportional to $\hat{Q}$ (not $\hat{A}$).
When considered from $\hat{Q}$ in addition to $\hat{P}$, WVs take on a  different perspective:
 in a sense, the entire $\hat{Q}$ space corresponds to an ideal measurement (i.e. $\Delta P$ small), because in a ideal (i.e. strong) measurement, $\hat{Q}$ is much broader.  However, if we start out as if we 
are going to do a strong measurement (i.e. $\Delta Q$ large), but then we post-select in just a part of MD, 
i.e. we look at just one small region of $\hat{Q}$, then  the whole structure of strong 
measurements can be understood as a quantum sums of WMs.  
Using eq. \ref{expweak}, Aharonov and Botero have shown \cite{ab} that different WVs are sampled as a result of the dispersion in $\hat{Q}$ and that an arbitrary strength ``ideal" measurement can be re-formulated as an average of WVs.  In \S \ref{rwm}, we implement these results for NSWM.

The new NSWM introduced in \S \ref{rwm} 
uses components of each of the 3 WM criterion and thus in the next 2 sections we review those aspects of SWM (in \S \ref{wvstat}) and STWM (in \S \ref{wvavgop}) most relevant to NSWM (SWM uses the first two criterion and 
STWM uses the third).

\label{errors}

\vskip -.5cm

\subsection{\bf  Statistical Weak Measurements (SWM)}

\label{wvstat}

With SWM, 
the WV is obtained robustly in a statistical sense 
from the mean reading of many separate pointers. I.e. an ensemble of $N$ separate systems and $N$ separate MDs are used. For {\bf each} individual system, in between it's pre- and post-selection, one of the $N$ MDs weakly measures the observable $\hat{A}$ of this single system and the outcome of this measurement is individually recorded. This is repeated for each of the $N$ different systems, each with a different MD. After the post-selection, the subset of those MDs that were associated with those systems which satisfied the post-selection criterion are collected out of the larger set of all possible post-selections (if all post-selections were included, then the decomposition, eq. \ref{expweak} would be reproduced). 
 A statistical analysis is then performed on the results of only those MDs associated with the proper post-selection. This reduces the uncertainty in the mean position  by $\frac{1}{\sqrt{N}}$,  thereby allowing for a more precise (i.e. robust) calculation of $\hat{A}_w$ ~\cite{av}.

We consider an example that will be used in different contexts throughout this letter: suppose we prepare (i.e. pre-select) a spin-1/2 system with $\hat{\sigma}_x=+1\equiv\vert\!\!\uparrow_x\!\rangle $ at time $t_{\mathrm{in}}$ and  post-select $\hat{\sigma}_{\mathrm{y}}= +1\equiv\vert\!\!\uparrow_y\!\rangle$ at time $t_{\mathrm{fin}}$. 
 A non-trivial aspect of pre- and post-selection, i.e. WVs, can then be investigated by considering a WM at the intermediate time
$t$ of  the spin in a
direction $\hat{\xi}$ (e.g. at an angle of $45^{\circ}$ between $x$ and $y$):
$\hat{\sigma}_{\xi}=\hat{\sigma}_x \cos 45 +\hat{\sigma}_y\sin 45=\frac{\hat{\sigma}_x +\hat{\sigma}_y}{\sqrt{2}}$.
For a WM, the inhomogeneity in the magnetic field induces a shift in momentum that is less than the uncertainty 
$\delta P_\xi < \Delta P_\xi$ and thus a wave packet corresponding to $\frac{\hat{\sigma}_x+\hat{\sigma}_y}{\sqrt{2}}=1$ will be broadly overlapping with the wave packet corresponding to $\frac{\hat{\sigma}_x+\hat{\sigma}_y}{\sqrt{2}}=-1$ because the deflection $\delta l \propto \langle \vec{\sigma}\cdot \nabla  B\rangle$ will not be discernable from the noise, i.e. $\delta l\ll \Delta l$ (where $\Delta l$ is the dispersion in the particle beam). Thus, it cannot be determined whether any individual particle corresponds to $\frac{\hat{\sigma}_x+\hat{\sigma}_y}{\sqrt{2}}=\pm 1$.  
The center of this distribution must then be determined by using an ensemble of $N$ particles, where $N>\left\{\frac{\Delta P_{\xi}}{\delta P_{\xi}} \right\}^2$.  
With this pre- and post-selection,  approximately $N/2$ out of $N$ pre-selected particles will satisfy the post-selection criterion  
 and thus this result is not a rare outcome.  
When correlated with the post-selection, the measurement result (which was confirmed experimentally for an analogous observable, the polarization~\cite{RSH}) is:
$(\hat{\sigma}_{\xi=45^\circ})_{\mathrm{w}} =\frac{\langle \uparrow_y\vert( \hat{\sigma}_x +\hat{\sigma}_y)/\sqrt{2} 
\vert\!\uparrow_x
\rangle}{\langle{\uparrow_y}\vert{\uparrow_x}\rangle}= \sqrt{2}$.
For an individual spin, the component of spin
$\hat{\sigma}_{\hat{\xi}}$ is an eigenvalue, $\pm 1$, but the WV
$(\hat{\sigma}_{\hat{\xi}})_w=\sqrt{2}$ is $\sqrt{2}$ times
bigger,
 (i.e. lies outside the range of 
eigenvalues of ${\bf \hat{\sigma} \cdot n}$) and is thus called an ``eccentric weak value" (EWV).  
With the SWM approach, each MD is quite
imprecise and we could not say whether the
distribution of results in the pointer was due to the original uncertainty in the pointer $\lambda \Delta Q$,
or due to the distribution of the observable of the system being measured, $\Delta A$.  
Therefore, a robust determination of the WV was obtained in a statistical sense: $N$ separate measurements were performed and recorded and the average was manually calculated. 
It is this aspect that is altered by NSWM.  t-selection.  

\vskip -.5cm

\subsection{\bf Single Trial Weak Measurements (STWM)}
\label{wvavgop}

Another aspect to obtaining NSWMs and shifting the statistical interpretation of WVs is the ability to measure {\it collective observables} on 
physical ensembles of $N$ particles.  As will be seen in this section,  this
allows us to measure all WVs with great precision
in one single (though previously thought to be rare) experiment.   
We first consider the following theorem:
\beq\label{identity}
\hat{A} |\Psi \rangle = \bar{A} \ket{\Psi}  + \Delta A \ket{\Psi_\perp}\, ,
\label{thm1}
\eeq
where $\bar{A} = \ave{\Psi}{\hat{A}}$, $|\Psi\rangle$ is any vector in Hilbert space, $\Delta A^2 = \ave{\Psi}{(\hat{A} - \bar{A})^2}$
and $\ket{\Psi_\perp}$ is some state such that $\amp {\Psi}{\Psi_\perp} = 0$.   Note that $\bar{A}$ is not defined here in a statistical sense: it is a mathematical property of an individual system $|\Psi\rangle$.  We can also {\it measure} this property with no reference to statistics by applying this identity to a composite, $N$-particle state
$\ket{\Psi\upn} =\ket{\Psi}_1\ket{\Psi}_2....\ket{\Psi}_N$
using 
a ``collective operator," $\hat{A}\upn\equiv \frac{1}{N} \sum_{\mathrm{i=1}}^{N} \hat{A}_i$ (where  $\hat{A}_i $ is the same operator $\hat{A}$ acting on the $i$-th particle).
Using this, we are able to obtain information on $\bar{A}$ without causing a collapse and thus without using a statistical approach 
because any product state
  $|\Psi \upn\rangle$ becomes an eigenstate of the operator $\hat{A}\upn$. 
To see this, consider $
{\hat{A}}\upn\ket{\Psi\upn}$:  
\beq
\hat{A}\upn\ket{\Psi\upn}  = \frac{1}{N}\left[ N \bar{A}\ket{\Psi\upn} + \Delta A \sum_i
|\Psi\upn_\perp(i) \rangle \right]
\label{avgop}
\eeq
where $\bar{A}$ is the average for any one particle and the states $|\Psi\upn_\perp(i) \rangle$ are mutually orthogonal and are given by
$|\Psi\upn_\perp(i) \rangle = \ket{\Psi}_1\ket{\Psi}_2...\ket{\Psi_\perp}_i...\ket{\Psi}_N$.
That is, the $i$th state has particle $i$ changed to an orthogonal state and all the other particles
remain in the same state.  If we further define a normalized state
 $|\Psi\upn_{\perp} \rangle = \sum_{i}\frac{1}{\sqrt{N}}|\Psi\upn_\perp(i) \rangle$ 
then the last term of eq. \ref{avgop} is $\frac{\Delta A}{\sqrt{N}}|\Psi\upn_{\perp} \rangle$ and it's size is $|\frac{\Delta A}{\sqrt{N}}
|\Psi\upn_{\perp} \rangle|^2\propto\frac{1}{N}\rightarrow 0$.
Therefore, $|\Psi\upn\rangle$ becomes an eigenstate of $\hat{A}\upn$, with value
$\overline{A}$, as $\hat{N} \rightarrow \infty$ (the second term decreases as $O(N^{-1/2})$ even if the particles are not all in the same state, as long as the composite
$N$-particle state is a product state).

To see the interplay between the pre- and post-selected  boundary conditions, we consider again a WM of the collective observable in the $45^{\circ}$ angle to the $x-y$ plane of $\hat{\sigma}_\xi\upn \equiv \frac{1}{N}
\sum_{\mathrm{i=1}}^{N} (\hat{\sigma}_\xi)_i$~\cite{aacv}. Using 
$H_{\mathrm{int}} = {{\lambda  (t)}\over N}  \hat{Q} \sum_{\mathrm{i=1}}^N  (\hat{\sigma}_i)_\xi$, a particular pre-selection of $|{\uparrow_x} \rangle$ (i.e. $|\Psi_{\mathrm{in}}\rangle = \prod_{\mathrm{i=1}}^N |{\uparrow_x} \rangle_i$) and post-selection

 $|{\uparrow_y}\rangle$
(i.e. 
$\langle\Psi_{\mathrm{fin}}|  = \prod_{\mathrm{i=1}}^N \langle{\uparrow_y}|_i=\prod_{\mathrm{n=1}}^N \left\{|{\uparrow_z} \rangle_n+i|{\downarrow_z}
\rangle_n\right\}$), then the pointer is robustly shifted by the
the same WV obtained in \S \ref{wvstat}, i.e. $\sqrt{2}$:
\beq
  \label{wvC}
  (\hat{\sigma}_\xi)_{\mathrm{w}} = {{\prod_{k=1}^N \langle{\uparrow_y}|_k ~ \sum_{\mathrm{i=1}}^N
\left\{(\hat{\sigma}_i)_x + (\hat{\sigma}_i)_y\right\} ~ \prod_{\mathrm{j=1}}^N |{\uparrow_x} \rangle_j}
\over { \sqrt 2 ~ N(\langle{\uparrow_y} |{\uparrow_x} \rangle)^N}}=
\sqrt2 \pm O(\frac{1}{\sqrt N}).
\label{wvlargespin}
\eeq
However, what is different from the SWM in \S \ref{wvstat} is that the imprecision of this measurement is $\pm \sqrt{N}$.
Thus, as $N \rightarrow \infty$ the intermediate WM will give $ (\hat{\sigma}_\xi)_{\mathrm{w}} = \weakv {\Psi_{\mathrm{fin}}}{ \hat{\sigma}_\xi}{\Psi_{\mathrm{in}}}\,=\sqrt{2}$ robustly since the shift in MD is greater than the uncertainty in MD.
A single experiment is now sufficient to
determine the WV with great precision and there is no longer
any need to average over results obtained in multiple experiments as we did in the previous section. 
Furthermore, if we repeat the experiment with different MDs, then each MD will
show the very same WVs, up to an insignificant spread of $\frac{1}{\sqrt
{ N}}$ (assuming we obtain the particular, rare, post-selection).  Therefore,  the information from {\it both} boundary conditions, i.e. 
$|\Psi_{\mathrm{in}}\rangle = \prod_{\mathrm{i=1}}^N |{\uparrow_x} \rangle_i$
 and 
$\langle\Psi_{\mathrm{fin}}|  = \prod_{\mathrm{i=1}}^N \langle{\uparrow_y}|_i$, describes the entire interval of time between pre- and post-selection (for plots see \cite{vaidman,Unruh2}).

STWM has 2 short-comings that are addressed by NSWM:
\begin{enumerate}

\item  while STWM is a valuable Gedanken experiment, 
the probability 
$|\langle\Psi_{\mathrm{fin}}|\Psi_{\mathrm{in}}\rangle|^{2N}$ for all $N$   particles to end up  in the same final state $|\Psi_{\mathrm{fin}} 
\rangle$ becomes exponentially small as  $N \rightarrow \infty$. \footnote{With the WM considered above, we have $N$ particles pre-selected with $\hat{\sigma}_x =1$, a WM of
$\hat{\sigma}_{45}$ (which
doesn't significantly disturb
the spins which are thus still in the state $\hat{\sigma}_x =1$ after the $\hat{\sigma}_{45}$ measurement) and followed by a post-selection in the y-direction.  The probability to obtain $\hat{\sigma}_y =1$ is $1/2$ and thus the total probability of finding all $N$
spins with $\hat{\sigma}_y =1$ is an exponentially small $2^{-N}$.}  Thus, with this method, the more accurately we measure $\hat{\sigma}_{45}$, the larger $N$ and thus the more rare the post-selection.

\item 
For STWM, the weakness of the interaction is built into the very collective observable being measured  and thus the effective coupling to each spin is smaller by a factor of $\frac{1}{N}$.  This keeps the interaction in the weak regime, even though the actual measurement strength $\lambda$ can be relatively large.  
However, for finite ensembles there is a limit to the size of $\lambda$,
and thus we cannot make the interaction stronger by a large factor.
\end{enumerate}

\vskip -.5cm

\section{\textcolor{black}{\bf  Non-statistical Weak Measurements (NSWM)}}
\label{rwm}

\vskip -.3cm

The NSWM introduced in this section shares positive attributes of both the ``statistical" (\S \ref{wvstat}) and ``single-trial" (\S \ref{wvavgop}) approaches
and 
allows us to shift the statistical meaning that was previously attributed to WMs  because:
\begin{enumerate}
\item 
We are only interested in  a single measurement (with an irreversible recording) of a robust quantity, the sum of momenta  $\sum_{\mathrm{i=1}}^{N} \hat{P}_{\mathrm{md}}^i$
 (not $N$ separate irreversible recordings of the momentum $\hat{P}_{\mathrm{md}}^i$ of each particle as was done in the ``statistical" WM in \S \ref{wvstat}).  
\item We are able to do this in a way that is not rare by using the approach in the first WM of \S \ref{wvstat}:  the easiest way
to pick out only those MDs that are associated with those systems that satisfied the proper post-selection criteria is to have  the particles themselves serve as MDs with the  information about  the WM stored, after the WM  interaction,  in a degree of  freedom separate from the pre- or post-selection, i.e. so that there is no coupling  between  the  variable  in which the result of the WM is stored and the  post-selection device. The post-selection of the particles then also selects out the relevant MDs. 
Therefore  the size of the ensemble required to obtain a precise WM is much smaller than in STWM.

\end{enumerate}

\begin{figure}[here]
\scalebox{.8}{\includegraphics{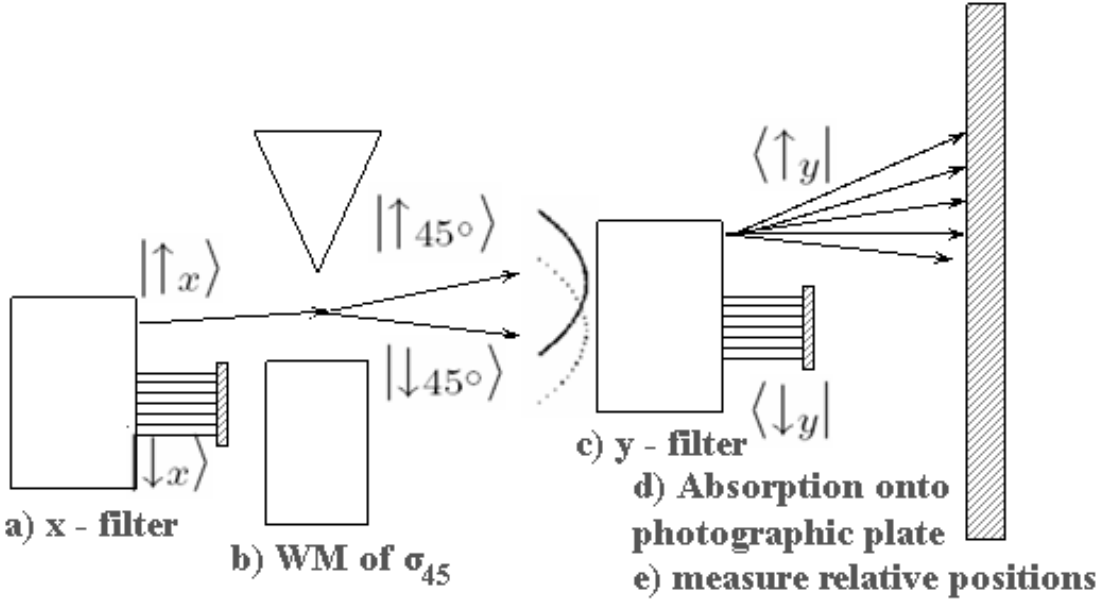}}
\caption[Stern Gerlach Apparatus]{NSWM by Stern-Gerlach apparatus for weakly measuring $\hat{\sigma}_{45}$.}
\label{sgpre}
%\vskip -3cm
\end{figure}

Measurement of the total momentum is an incomplete measurement and thus it's uncertainty will be larger than if it is conditioned on certain deviations of positions.
Therefore, after the measurement of the total momentum, we can use the relative positions to correct for the disturbance caused to the system which resulted from a stronger measurement interaction.  The disturbance due to larger strength measurements results in a different pre- or post-selection  which thereby translates to a different though still eccentric WV.
Without loss of generality, we present the new NSWM in the framework of a Stern-Gerlach Gedanken experiment used in previous sections:
consider a large collection of particles where the MD is simply the position and momentum of those particles (see fig. \ref{sgpre}).  We perform the following a) filter out $|{\downarrow_x} \rangle$ at time $t_{\mathrm{in}}$\footnote{The filter only interacts with the component of the spin that is not transmitted, e.g. the $\sigma_x=-1$ component would receive a strong repulsive interaction (via a potential $\hat{\sigma}_xB_x -B_x$ using a homogenous $B_x$), while the $\sigma_x=+1$ component would not have any change in it's momentum.}; b) perform a WM of $\frac{\hat{\sigma}_x+\hat{\sigma}_y}{\sqrt{2}}$ at the intermediate time $t$ but wait until after performing the post-selection to read out the result of the sum of $\approx \frac{N}{2}$ of these interactions\footnote{We recommend that a  WM of $\sigma_x$ uses a field $B_o(\sigma_x x-\sigma_y y)$ with a small width in $y$.  There is then little variation in the wavefunction in the y-direction.  The x-direction  would not be constrained and the wavefunction can freely vary in $x$.  With this method, only a shift in the $x$ direction would occur; the wavefunction in $y$ would always be left in the ground state because the force is too small to excite it.}; c) filter out  $\langle{\downarrow_y}|$ at time $t_{\mathrm{fin}}$; 
d) absorb the particles onto a photographic plate 
and measure the sum of momenta  $\sum_{\mathrm{i=1}}^{N} \hat{P}_{\mathrm{md}}^i$
(without measuring the individual $\hat{P}_{\mathrm{md}}^i$); this recording  will produce a definite shift 
by a WV; e) measure the relative positions to determine what the pre-selected and post-selected system the WM in step (b) was a measurement of. 
\vskip -.3cm
%\vskip -.3cm

Let us consider again 
a particular pre-selection for each particle of $|{\uparrow_x} \rangle$ and post-selection in the
state $|{\uparrow_y}\rangle$, so $|\Psi_{\mathrm{in}}\rangle = \prod_{\mathrm{j=1}}^N |{\uparrow_x} \rangle_j$ 
 and  $|\Psi_{\mathrm{fin}}\rangle  = \prod_{\mathrm{j=1}}^N |{\uparrow_y}\rangle_j$. 
The WM interaction in step b)~\cite{aacv} is  described by an interaction Hamiltonian which couples $\hat{\sigma}_\xi$ with $\hat{Q}_{\mathrm{md}}$ of MD
i.e. $H_{\mathrm{int}} = \lambda  (t)   \hat{Q}_{\mathrm{md}}^j\hat{\sigma}^j_\xi$ (where $\hat{\sigma}^j_\xi=\{ \frac{\hat{\sigma}_x+\hat{\sigma}_y}{\sqrt{2}}\}_j$).
This generates shifts in the individual momenta due to the WM interaction (as occurred in \S \ref{wvstat}).  However, unlike \S \ref{der2v} there is only {\bf one irreversible recording} of the sum of these shifts (as occurred in \S \ref{wvavgop}), i.e. one irreversible recording of the total momentum of the $N$ particles which were deposited onto a single photographic plate, followed by $N$ measurements of the  positions (used to deduce the $N-1$ relative positions).  
This state of the photographic plate after its interaction with the $N$ particles shall be referred to as the ``final state of the MD".  

 After the WM interaction and post-selection  (but before the irreversible recording), 
the final state of MD is:
\beq
\Phi_{\mathrm{fin}}^{\mathrm{MD}}=\prod_{\mathrm{j=1}}^N \langle{\uparrow_y}|_j \exp \lbrace {\mathrm{i}}\lambda \hat{Q}_{\mathrm{md}}^j \hat{\sigma}^j_\xi\rbrace  |{\uparrow_x} \rangle_j \exp\left\{- {({{Q}_{\mathrm{md}}^j)^2}\over{4(\Delta {Q}_{\mathrm{md}}^j)^2}}\right\}
\label{bigspin}
\eeq
Here we have set the coupling to each spin to be $ \int \lambda  (t)dt=\lambda $
and without loss of generality have taken the initial state of MD as simply a Gaussian in the coordinate $\hat{Q}_{\mathrm{md}}^j$ of each particle, i.e. $(\Phi_{\mathrm{in}}^{\mathrm{MD}})^j=\exp\left\{- {({{Q}_{\mathrm{md}}^j)^2}\over{4(\Delta {Q}_{\mathrm{md}}^j)^2}}\right\}$.

As will be seen later, it will prove useful to reformulate the MD observables in terms of two complementary, non-commuting, collective observables, a pointer corresponding to the sum of momenta, $\hat{P}_{\mathrm{md}}^{\mathrm{(N)}}$, and it's conjugate $\hat{Q}_{\mathrm{md}}^{\mathrm{(N)}}$ which generates shifts in the pointer $\hat{P}_{\mathrm{md}}^{\mathrm{(N)}}$: 
\begin{eqnarray}
\hat{P}_{\mathrm{md}}^{\mathrm{(N)}}&\equiv& \sum_{\mathrm{j=1}}^{N} \frac{\hat{P}_{\mathrm{md}}^j}{\sqrt{N}}\\
\hat{Q}_{\mathrm{md}}^{\mathrm{(N)}}&\equiv& \sum_{\mathrm{j=1}}^N \frac{\hat{Q}_{\mathrm{md}}^j}{\sqrt{N}}
\end{eqnarray}
  These definitions are particularly useful because
if the uncertainty in the individual $\hat{Q}_{\mathrm{md}}^j$'s is $\Delta \hat{Q}_{\mathrm{md}}^j\sim 1$, then the uncertainty in $\hat{Q}_{\mathrm{md}}^{\mathrm{(N)}}$ is also $\Delta \hat{Q}_{\mathrm{md}}^{\mathrm{(N)}}\sim 1$ due to $[\hat{Q}_{\mathrm{md}}^{\mathrm{(N)}}, \hat{P}_{\mathrm{md}}^{\mathrm{(N)}}]=1$ (The spread in $\sum_{\mathrm{j=1}}^N \hat{Q}_{\mathrm{md}}^j$ is $\sqrt{N}$ (as occurs in a Poisson distribution) and thus $\frac{\sum_{\mathrm{j=1}}^N \hat{Q}_{\mathrm{md}}^j}{\sqrt{N}}\approx 1$).  Using

\begin{eqnarray}
- \sum_{\mathrm{j=1}}^N\frac{({Q}_{\mathrm{md}}^j)^2}{4(\Delta {Q}_{\mathrm{md}}^j)^2}&=&- \frac{1}{4(\Delta {Q}_{\mathrm{md}}^j)^2}\left[{\sum_{\mathrm{j=1}}^N{\lbrace{{Q}_{\mathrm{md}}^j}}- \frac{{Q}_{\mathrm{md}}^{\mathrm{(N)}}}{\sqrt{N}}\rbrace^2}+\sum_{\mathrm{j=1}}^N\lbrace  2{{Q}_{\mathrm{md}}^j}\frac{{Q}_{\mathrm{md}}^{\mathrm{(N)}}}{\sqrt{N}}-\frac{[{Q}_{\mathrm{md}}^{\mathrm{(N)}}]^2}{N}\rbrace\right]\nonumber\\
&=&- \frac{1}{4(\Delta {Q}_{\mathrm{md}}^j)^2}{\sum_{\mathrm{j=1}}^N{\lbrace{{Q}_{\mathrm{md}}^j}}- \frac{{Q}_{\mathrm{md}}^{\mathrm{(N)}}}{\sqrt{N}}\rbrace^2}
+[{Q}_{\mathrm{md}}^{\mathrm{(N)}}]^2
\end{eqnarray}
it will also be useful to re-write the wavefunction of MD as:
\beq
\prod_{\mathrm{j=1}}^N\exp\left\{- {({{Q}_{\mathrm{md}}^j)^2}\over{4(\Delta {Q}_{\mathrm{md}}^j)^2}}\right\}\rightarrow\exp{ \lbrace-{{({Q}_{\mathrm{md}}^{\mathrm{(N)}})^2}\over{4(\Delta {Q}_{\mathrm{md}}^{\mathrm{(N)}})^2}}\rbrace}\exp{ \lbrace-{{\sum_j({Q}_{\mathrm{md}}^j-\frac{{Q}_{\mathrm{md}}^{\mathrm{(N)}}}{\sqrt{N}})^2}\over{4(\Delta {Q}_{\mathrm{md}}^{\mathrm{(N)}})^2}}\rbrace}
\label{wfmd}
\eeq

We now show that measuring the relative positions provides corrections to the pre- or post-selection, thus giving a different WV for each particle, represented by $\tilde{\sigma}_{\mathrm{w}}^j$ (which thereby explains the utility of $\exp{ \lbrace-{{\sum_j({Q}_{\mathrm{md}}^j-\frac{{Q}_{\mathrm{md}}^{\mathrm{(N)}}}{\sqrt{N}})^2}\over{4(\Delta {Q}_{\mathrm{md}}^{\mathrm{(N)}})^2}}\rbrace}$).  

\vskip -.5cm

\subsection{\bf Use of relative positions}

\label{relposcorrect}

Besides the sum of momenta, we can also measure  the  $N-1$ relative positions (without disturbing the system),
\beq
\hat{x}_i=\hat{Q}_{\mathrm{md}}^i-\sum \frac{\hat{Q}_{\mathrm{md}}^n}{N}=\hat{Q}_{\mathrm{md}}^i-\frac{\hat{Q}_{\mathrm{md}}^{\mathrm{(N)}}}{\sqrt{N}}
\eeq
 This is because 
$[\hat{Q}_{\mathrm{md}}^i-\sum \frac{\hat{Q}_{\mathrm{md}}^n}{N},\sum \hat{P}_{\mathrm{md}}^i]=[\hat{x}_i,\hat{P}_{\mathrm{md}}^{\mathrm{(N)}}]=0$
which is easy to see because each pair of relative positions commutes with the sum of momenta, i.e. $[\hat{Q}_{\mathrm{md}}^i- \hat{Q}_{\mathrm{md}}^j,\sum_{\mathrm{n=1}}^{N} \hat{P}_{\mathrm{md}}^n]=[\hat{Q}_{\mathrm{md}}^i,\sum_{\mathrm{n=1}}^{N} \hat{P}_{\mathrm{md}}^n]-[\hat{Q}_{\mathrm{md}}^j,\sum_{\mathrm{n=1}}^{N} \hat{P}_{\mathrm{md}}^n]=i\hbar-i\hbar=0$, using $[\hat{Q}_{\mathrm{md}}^j,\sum_{\mathrm{n=1}}^{N} \hat{P}_{\mathrm{md}}^n]=[\hat{Q}_{\mathrm{md}}^j, \hat{P}_{\mathrm{md}}^j]=i\hbar$.  Furthermore $\hat{Q}_{\mathrm{md}}^i-\sum_{\mathrm{n=1}}^{N} \frac{\hat{Q}_{\mathrm{md}}^n}{N}=\sum_{\mathrm{n=1}}^{N} \frac{\hat{Q}_{\mathrm{md}}^i-\hat{Q}_{\mathrm{md}}^n}{N}$.

As a preparation to obtain both a measurement of the relative positions and of the total momenta  we re-write eq. \ref{bigspin} as: 
\begin{eqnarray}
\Phi_{\mathrm{fin}}^{\mathrm{MD}}&=& \prod_{\mathrm{j=1}}^N \langle{\uparrow_y}|_j \exp \lbrace {\mathrm{i}}\lambda\underbrace{\{ \hat{Q}_{\mathrm{md}}^j-\sum_{\mathrm{n=1}}^N \frac{\hat{Q}_{\mathrm{md}}^n}{N}\}}_{\mathrm{relative\, positions}} \hat{\sigma}^j_\xi\rbrace \underbrace{\exp \lbrace {\mathrm{i}}\lambda  \sum_{\mathrm{n=1}}^N \frac{\hat{Q}_{\mathrm{md}}^n}{N} \hat{\sigma}^j_\xi \rbrace }_{\mathrm{induces \, translations \, in \, \hat{Q}_{\mathrm{md}}^{\mathrm{(N)}}}}{|\uparrow_x} \rangle_j\nonumber\\
&\bigotimes & \exp{ \lbrace-{{({Q}_{\mathrm{md}}^{\mathrm{(N)}})^2}\over{4(\Delta {Q}_{\mathrm{md}}^{\mathrm{(N)}})^2}}\rbrace}\exp{ \lbrace-{{
\sum_j({Q}_{\mathrm{md}}^j-\frac{{Q}_{\mathrm{md}}^{\mathrm{(N)}}}{\sqrt{N}})^2}\over{4(\Delta {Q}_{\mathrm{md}}^{\mathrm{(N)}})^2}}\rbrace}\nonumber\\
&=&\prod_{\mathrm{j=1}}^N \langle{\uparrow_y}|_j \exp \lbrace {\mathrm{i}}\lambda \hat{x}_j \hat{\sigma}^j_\xi\rbrace \exp \lbrace {\mathrm{i}}
\lambda\frac{\hat{Q}_{\mathrm{md}}^{\mathrm{(N)}}}{\sqrt{N}} \hat{\sigma}^j_\xi \rbrace |{\uparrow_x} \rangle_j\nonumber\\
&\bigotimes & \exp{ \lbrace-{{({Q}_{\mathrm{md}}^{\mathrm{(N)}})^2}\over{4(\Delta {Q}_{\mathrm{md}}^{\mathrm{(N)}})^2}}\rbrace}\exp{ \lbrace-{{\sum_j({Q}_{\mathrm{md}}^j-\frac{{Q}_{\mathrm{md}}^{\mathrm{(N)}}}{\sqrt{N}})^2}\over{4(\Delta {Q}_{\mathrm{md}}^{\mathrm{(N)}})^2}}\rbrace}
\label{bigspin2}
\end{eqnarray}

How is this re-formulation of eq. \ref{bigspin} in terms of the relative positions $\hat{x}_j=\hat{Q}_{\mathrm{md}}^j-\sum \frac{\hat{Q}_{\mathrm{md}}^n}{N}$  helpful?  To see this, we'll consider eq. \ref{bigspin2} one particle at a time.  For the $jth$ particle, we can apply the first exponential of eq. \ref{bigspin2}, $\exp \lbrace i\lambda \hat{x}_j \hat{\sigma}^j_\xi\rbrace$, to either the pre-selected state $|{\uparrow_x} \rangle_j$  or to the post-selected state $\langle{\uparrow_y}|_j$ (since the 2 exponentials commute).  
What does this exponential do to   the pre- or post-selection? 
As  mentioned in \S \ref{der2v},
$\hat{Q}_{\mathrm{md}}^{\mathrm{(N)}}$ and $\hat{A}$ (or in this case $\hat{x}_j$ and $\hat{\sigma}^j_\xi$) have 2 inverse roles:  the back reaction on the {\it system} is generated by $\hat{\sigma}^j_\xi$ in a manner proportional to $\hat{x}_j$. 
However, the $\hat{x}_j$ can be measured exactly and can thus be replaced by a number.  Therefore,  $\exp \lbrace i\lambda \hat{x}_j \hat{\sigma}^j_\xi\rbrace$ simply 
 rotates the pre- or post-selected state about the axis $\xi $ by an angle given by $\lambda x_j$:
\beq
\underbrace{\exp \lbrace {\mathrm{i}}\lambda \hat{x}_j 
\hat{\sigma}^j_\xi\rbrace}_{\mathrm{rotates \,\, bra \,\, by \,\, a\,\, definite\,\, amount}} |{\uparrow_x} \rangle_j\equiv |\Psi \rangle_j
\label{bigspin4}
\eeq
Thus, measurement of the relative positions allows us to definitely determine how much $\exp \lbrace {\mathrm{i}}\lambda \hat{x}_j 
\hat{\sigma}^j_\xi\rbrace$ rotates $|{\uparrow_x} \rangle_j$ (i.e. to $|\Psi \rangle_j$). 
Therefore, eq. \ref{bigspin4} acts as a correction to the ensemble:
instead of the original ensemble of pre-selected $|{\uparrow_x} \rangle$ and post-selected $\langle{\uparrow_y}|$ states, we will have a new  ensemble with  shifted pre- or post-selections. 

How could the relative positions be measured?  Procedurally, we first measure the momentum of the photographic plate after the $N$ particles have deposited their momentum.
When we subtract from this the initial  momentum of the photographic plate, then we can determine the shift in the sum of the momentum for the $N$ particles as a result of the WM interaction in a new way.  After the final measurement of $\hat{P}_{\mathrm{md}}^{\mathrm{(N)}}$, 
we then measure the $N$ individual positions (i.e. $\hat{Q}_{\mathrm{md}}^j$) of each particle that is deposited onto the photographic.
Now, measurement of $\hat{P}_{\mathrm{md}}^{\mathrm{(N)}}$
will disturb the individual $\hat{Q}_{\mathrm{md}}^j$'s but will not disturb the relative positions (since they commute with the total momenta).  
Therefore, even though the subsequent measurement of the $N$ $\hat{Q}_{\mathrm{md}}^j$'s will be un-related to the value of the $\hat{Q}_{\mathrm{md}}^j$'s during the WV,  
 we can {\it deduce} what the relative positions were at the time of the WM through the individual positions.
\footnote{
If the uncertainty in the individual $\hat{Q}_{\mathrm{md}}$'s is $\Delta \hat{Q}_{\mathrm{md}}\sim 1$, then the uncertainty in $\hat{Q}_{\mathrm{md}}^{\mathrm{(N)}}$ is also $\Delta Q\sim 1$ (because the spread in $\sum\hat{Q}_{\mathrm{md}}$ is $\sqrt{N}$ and thus $\frac{\sum\hat{Q}_{\mathrm{md}}}{\sqrt{N}}\approx 1$).  Therefore, the spread in $\frac{\sum\hat{Q}_{\mathrm{md}}}{N}$ is negligible and thus $\hat{x}_i=\hat{Q}_{\mathrm{md}}^i-\sum \frac{\hat{Q}_{\mathrm{md}}^n}{N}$ also has the same uncertainty as $\Delta \hat{Q}_{\mathrm{md}}$.}

After substituting the single particle result 
eq. \ref{bigspin4}  for the $jth$ particle (i.e. using the rotated bra or ket), back into the $N$ particle eq. \ref{bigspin2}, we have:
\beq
\prod_{\mathrm{j=1}}^N \langle{\uparrow_y}|_j  \exp \lbrace {\mathrm{i}}\frac{\lambda}{\sqrt{N}}{\hat{Q}_{\mathrm{md}}^{\mathrm{(N)}}} \hat{\sigma}^j_\xi \rbrace |\Psi\rangle_j\exp\left\{-{{({Q}_{\mathrm{md}}^{\mathrm{(N)}})^2}\over{4(\Delta {Q}_{\mathrm{md}}^{\mathrm{(N)}})^2}}\right\}
\label{bigspin6}
\eeq
It is clear from eq. \ref{bigspin6} that when we look at the particles that are left unknown after using all the information (both relative positions and the total momenta) and consider them as the final total spin, then it is like a robust experiment but now the coupling to each spin is $\frac{\lambda}{\sqrt{N}}$ 
and thus the remaining effect of the coupling in the exponential will be small.

\vskip -.5cm

\subsection{\bf Proving the legitimacy of WVs for a new regime}
\label{newway}

After using these corrections, we can now prove the validity of the WV approximation. We will show that the final state of MD, i.e. of eq. \ref{bigspin6}  will be:

\beq
\Phi_{\mathrm{fin}}^{\mathrm{MD}}= \exp \lbrace \frac{{\mathrm{i}}\lambda \hat{Q}_{\mathrm{md}}^{\mathrm{(N)}}}{\sqrt{N}} \sum_{\mathrm{j=1}}^N \tilde{\sigma}_{\mathrm{w}}^j  \rbrace\exp\lbrace-{{({Q}_{\mathrm{md}}^{\mathrm{(N)}})^2}\over{4(\Delta {Q}_{\mathrm{md}}^{\mathrm{(N)}})^2}}\rbrace
\label{finalwm}
\eeq
($\tilde{\sigma}_{\mathrm{w}}^j$ is the WV for the $jth$ particle
- a tilde will always refer to WVs calculated with rotated states)
When this is transformed back to the momentum 
representation (as was done in the WV approximation used in eq. \ref{mstate2}), then the momentum of MD is shifted by the WV; i.e. the change in $\hat{P}_{\mathrm{md}}^{\mathrm{(N)}}$ (the change in the sum of momentum $\sum_{\mathrm{i=1}}^{N}\hat{P}_{\mathrm{md}}^i$) is:
\beq
\delta \hat{P}_{\mathrm{md}}^{\mathrm{(N)}}=\delta \sum_{\mathrm{i=1}}^{N} \frac{\hat{P}_{\mathrm{md}}^i}{\sqrt{N}} = \frac{\lambda}{\sqrt{N}}\sum_{\mathrm{j=1}}^N \tilde{\sigma}_{\mathrm{w}}^j
\label{finprf}
\eeq

To prove the legitimacy of this WV calculation,  
we first assume for simplicity a small variance in the rotations so that each particle yields approximately the same WV, i.e. $\tilde{\sigma}_{\mathrm{w}}^j\equiv\bar{\alpha}_{\mathrm{w}}$, enabling us to re-write eq. \ref{finalwm} as:
\beq
\{cos \frac{\lambda \hat{Q}_{\mathrm{md}}^{\mathrm{(N)}}}{\sqrt{N}} +{\mathrm{i}}\bar{\alpha}_{\mathrm{w}} sin \frac{\lambda \hat{Q}_{\mathrm{md}}^{\mathrm{(N)}}}{\sqrt{N}}\}^N e^{-\frac{({Q}_{\mathrm{md}}^{\mathrm{(N)}})^2}{4(\Delta {Q}_{\mathrm{md}}^{\mathrm{(N)}})^2}}
\label{bigspin7}
\eeq
Now, in order to perform a valid WV calculation, this function needs to be peaked around $\hat{Q}_{\mathrm{md}}^{\mathrm{(N)}}=0$. As long as there are no regions in which the size of eq. \ref{bigspin7} (i.e. eq. \ref{bigspin8}) exceed the exponential of MD then it will be as if we are around $\hat{Q}_{\mathrm{md}}^{\mathrm{(N)}}=0$.   In other words, the legitimacy of the WV calculation can now be understood as a competition between the (scalar product) $A$ term and the (probability) $B$ term: 
\beq
|\Phi_{\mathrm{fin}}^{\mathrm{MD}}|=\underbrace{\{1 + (\bar{\alpha}_{\mathrm{w}}^2 -1)sin^2 \frac{\lambda \hat{Q}_{\mathrm{md}}^{\mathrm{(N)}}}{\sqrt{N}}\}^{\frac{N}{2}}}_A \underbrace{\exp\left\{-\frac{({Q}_{\mathrm{md}}^{\mathrm{(N)}})^2}{2(\Delta {Q}_{\mathrm{md}}^{\mathrm{(N)}})^2}\right\}}_B
\label{bigspin8}
\eeq
If  the quantity eq. \ref{bigspin8} goes to $0$ for large $Q$ then the WV approximation is valid.
On the other hand, if the increase in  $A$ was not counter-balanced by the decline in $B$ then we could not restrict the WV approximation around $Q=0$ because it would be much more likely to be located around large $Q$.   Thus, the meaning of the new WV approximation presented here is that there is no other region in which the size of eq. \ref{bigspin7}, i.e. eq. \ref{bigspin8}, is significant, except around $Q=0$.

We now  ask what is the maximum value of $\lambda $ such that we still obtain a shift in the pointer by $\frac{\lambda}{\sqrt{N}}\sum_{\mathrm{j=1}}^N \bar{\alpha}_{\mathrm{w}}$? 
We will see that the constraint ${\lambda}  \ll 1$ as was required in \S \ref{wvstat} (and in the first article on WMs~\cite{aacv}) is unneccessary in order to obtain a valid WM.  In fact, with NSWM the coupling to each individual spin just needs to be $\lambda \sim 1$.

\vskip -3cm

\subsubsection{\bf  N$\rightarrow \infty$, $Q$ finite:}

First we consider 
how large $\lambda$ can be for a legitimate WV in the regime $N\rightarrow 
\infty$ with $Q$ finite.  In this case, we consider again the magnitude 
(eq. \ref{bigspin8}).  $A$ can be written as $\lbrace 1+(\bar{\alpha}_{\mathrm{w}}^2-1)
\left[\frac{\lambda^2 Q^2}{N}- \frac{\lambda^4 Q^4}{3N^2}\right]\rbrace^{
\frac{N}{2}}$.  If $Q$ is finite when $N\rightarrow \infty$ then we can 
neglect  $\frac{1}{N^2}$ and higher terms from the expansion of $\sin^2$.  
Thus $A\approx \lbrace 1+(\bar{\alpha}_{\mathrm{w}}^2-1)\frac{\lambda^2 Q^2}{N}\rbrace^{
\frac{N}{2}}\approx\exp\lbrace \frac{(\bar{\alpha}_{\mathrm{w}}^2-1)\lambda^2 Q^2}{2}\rbrace$. 
 As long as $(\bar{\alpha}_{\mathrm{w}}^2-1)\lambda^2 < 1$, then the increase in  $A$, i.e.  $\exp \lbrace \frac{{\mathrm{i}}\lambda \hat{Q}_{\mathrm{md}}^{\mathrm{(N)}}}{\sqrt{N}} \sum_{\mathrm{j=1}}^N \tilde{\sigma}_{\mathrm{w}}^j  \rbrace$, is counter-balanced by the decline in the Gaussian $B$, and  thus eq. \ref{bigspin8} is centered around $Q=0$.

Now $\sum_{\mathrm{j=1}}^N \bar{\alpha}_{\mathrm{w}}\approx N \bar{\alpha}_{\mathrm{w}}$ and thus 
all $N$ particles will deliver a momentum shift to the photographic plate of $\lambda N\bar{\alpha}_{\mathrm{w}}$ (and a shift in $\hat{P}_{\mathrm{md}}^{\mathrm{(N)}}$ of  $\lambda \sqrt{N}\bar{\alpha}_{\mathrm{w}}$ ).  The shift goes up as $N\bar{\alpha}_{\mathrm{w}}$  while the uncertainty goes up as ${\sqrt{N}}$ (the variance is 
$\langle (\Delta \hat{P}_{\mathrm{md}}^{\mathrm{(N)}})^2\rangle=[(\Delta \hat{P}_{\mathrm{md}})^2+\langle(\Delta (\sigma_\xi)_{\mathrm{w}})^2\rangle]$) and thus $\frac{\langle \hat{P}_{\mathrm{md}}^{\mathrm{(N)}}\rangle}{\langle \Delta \hat{P}_{\mathrm{md}}^{\mathrm{(N)}}\rangle}\approx\sqrt{N}$.   By choosing a sufficiently large $N$, the single trial WM outcome can be arbitrarily amplified. 

We have thus shown that it is valid to perform a WV approximation (i.e. to replace eq. \ref{bigspin6} with eq. \ref{finalwm}) in a significantly stronger coupling regime, i.e. for $\lambda \sim 1$,
by measuring a variable where it's shift is large compared to it's noise and thus, this is a NSWM.  
\label{finiteq}

\vskip -.5cm

\subsubsection{\bf Finite N} 
\begin{figure}[here] 
{\includegraphics{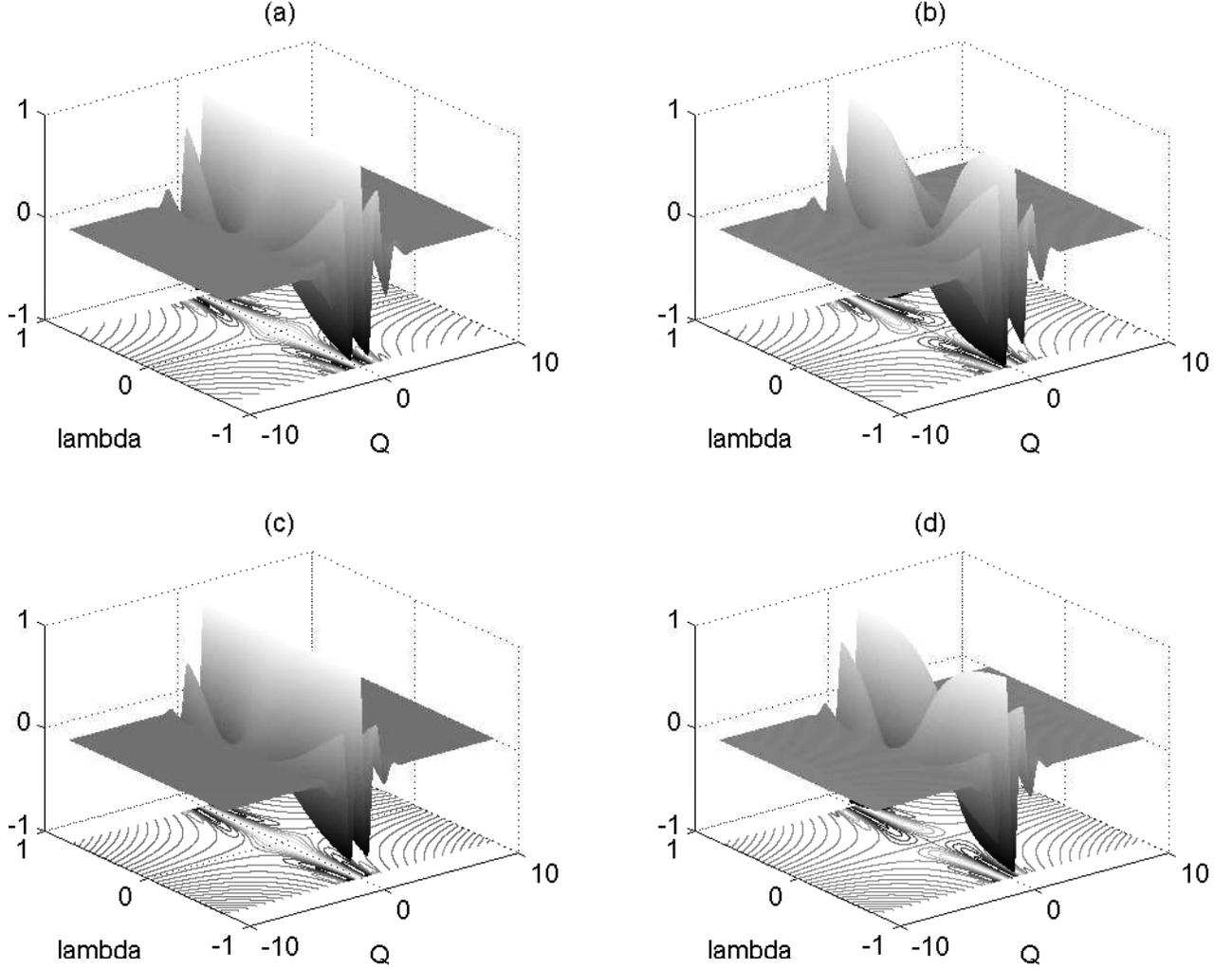}} 
\caption[Stern Gerlach Apparatus]{Numerical comparison of the NSWM wavefunction (a-b) with the ideal shift by the WV $(\sigma_{45})_w=\sqrt{2}$ (c-d) for $N=20$. a) Real part of eq. \ref{bigspin6}, b) Imaginary part of eq. \ref{bigspin6}, c) Real part of eq. \ref{finalwm}, d) Imaginary part of eq. \ref{finalwm}.}\label{sgpre8} \end{figure}

We consider finite $N$ where there is a proper limit in which $N$ increases and the interaction goes to $0$ but $\lambda ^3$ is negligible.  If we fix $\lambda$ and choose an $N$ such that $\lambda \sqrt{N} > 1$, then we can measure the average exactly.  The uncertainty of $P$ for $N$ particles is $\sqrt{N}$ and the momentum grows as $\lambda N \bar{\alpha} > \sqrt{N}$ which implies that $2\sqrt{N} \bar{\alpha} > 1$.  Nevertheless $N\lambda^3$ is still small (i.e. $N\lambda> \sqrt{N}$) but $N\lambda^3 < \frac{1}{\sqrt{N}}$ and $N\lambda^3$ is the extra correction.
For each spin there is a correction proportional to $\lambda ^3$ which for $N$ particles is $N\lambda^3$ which is small compared to $\sqrt{N}$ so $\sqrt{N} \lambda^3 < 0$ can be neglected.
We plot $N=20$ (fig. \ref{sgpre8}) to show that eq. \ref{finalwm}
 is an accurate approximation to eq. \ref{bigspin6}.

\vskip -3cm

\subsubsection{\bf  Quantum Average of WVs:}

In the last 2 regimes, 
we chose, for simplicity, to ignore the details of a significant variation in the WVs, 
e.g. $\sum_{\mathrm{j=1}}^N \tilde{\sigma}_{\mathrm{w}}^j\approx N\bar{\alpha}_{\mathrm{w}}$.  Normally, the WV is not very sensitive to different pre- or post-selections: i.e. if 2 possible pre- or post-selections are close to each other, i.e. $\ket{\Psi_{\mathrm{in}}}\approx\ket{\tilde{\Psi}_{\mathrm{in}}}$ then the WVs, $\weakv {\Psi_{\mathrm{fin}}}{\hat{A}}{\Psi_{\mathrm{in}} }$ and $\weakv {\Psi_{\mathrm{fin}}}{\hat{A}}{\tilde{\Psi}_{\mathrm{in}} }$ are also close.  However, in the instant case, the shift generated in the pointer 
can be large, so even a small rotation to the pre- or post-selection will make a significant difference in the pointer shift.
Even if the variance were significant, it is easy to see that our result is still valid.  I.e. even if the individual $\Psi$'s in the composite state are very different, then eq. \ref{avgop} is still valid, only the average will be over different pre- and post-selections.
 While it is appropriate to replace $\sigma$ by it's WV, 
there are 2 reasons that rotations can be induced in the pre- and/or post-selection.
Up to this point we have only discussed the first rotation (e.g. from $|{\uparrow_x} \rangle_j$  to $|\Psi \rangle_j$) which 
was corrected by the measurement of the relative positions.  This produced a shift in the pointer by an average over well-known WVs.
  However, $\exp \lbrace \frac{i\lambda \hat{Q}_{\mathrm{md}}^{\mathrm{(N)}}}{\sqrt{N}} \sum_{\mathrm{j=1}}^N \tilde{\sigma}_{\mathrm{w}}^j  \rbrace$ can cause a second rotation in the pre- or post-selection if  $\lambda\Delta Q$ given by eq. \ref{bigspin7} is big enough.
We now analyze this  {\it additional} rotation in the pre- or post-selections  which cannot be determined by measurement of the relative positions and show
that the total momentum, $\hat{P}_{\mathrm{md}}^{\mathrm{(N)}}$, is shifted by a quantum average of WVs~\cite{ab}
with weights determined by the probability to obtain a particular $Q$ which is associated with a particular WV as suggested by eq. \ref{expweak}.  
As a simple example, one may  categorize the different WVs into different pre- or post-selections.  Suppose a subset, $n_1$, out of the ensemble of $N$ particles will all be rotated to the same state (e.g. to $|\Psi_1\rangle$) and thus will give one WV $\tilde{\eta}_{\mathrm{w}}^1$, other subsets will be rotated to another state (e.g. to $|\Psi_2\rangle$) giving another WV $\tilde{\eta}_{\mathrm{w}}^2$, etc.
Using $\sum n_i=N$.\footnote{For any product state, we still have that eq. \ref{thm1} is exactly true but with an $\bar{\sigma}=\sum {\tilde{\eta}}_n$, i.e. $\hat{Q}_{\mathrm{md}}^{\mathrm{(N)}}\hat{\sigma}\upn\ket{\Psi\upn}  = \hat{Q}_{\mathrm{md}}^{\mathrm{(N)}}\sum \tilde{\eta}_n\ket{\Psi\upn} + \hat{Q}_{\mathrm{md}}^{\mathrm{(N)}}\frac{\Delta
\sigma}{\sqrt{N}} |\Psi\upn\perp \rangle$.}
eq. \ref{bigspin6} is re-written:
\beq
\left[\exp \{ {\mathrm{i}}\lambda \hat{Q}_{\mathrm{md}}^{\mathrm{(N)}}\frac{n_1}{\sqrt{N}}\tilde{\eta}_{\mathrm{w}}^1\} 
\cdot\cdot\cdot\exp \{ {\mathrm{i}}\lambda \hat{Q}_{\mathrm{md}}^{\mathrm{(N)}}\frac{n_k}{\sqrt{N}}\tilde{\eta}_{\mathrm{w}}^n\}\right]\exp\left\{-{{({Q}_{\mathrm{md}}^{\mathrm{(N)}})^2}\over{4(\Delta {Q}_{\mathrm{md}}^{\mathrm{(N)}})^2}}\right\}
\label{toprove}
\eeq

Normally a valid WV calculation requires MD to be centered around $\hat{Q}_{\mathrm{md}}^{\mathrm{(N)}}=0$.  However, in \cite{ab} ideal measurements were converted to WMs  by post-selecting MD to be in a certain region of $Q$ and in $P$, i.e. different regions of $\hat{Q}$ were sampled by multiplying by a function centered at $Q_{\mathrm{com}}$: i.e. a function of $Q'=Q-Q_{\mathrm{com}}$ such as  $\exp \frac{-(Q-Q_{\mathrm{com}})^2}{\Delta Q^2}$ which is like starting the MD not with $Q=0$ but with $Q=Q_{\mathrm{com}}$.  Results centered at different $Q=Q_{\mathrm{com}}$ are  then summed.   However, even such limited projections can still disturb each other.   The new NSWM presented here is more subtle because the relative coordinates commute with the total momentum and so can be simultaneously measured without disturbing each other.  
By measuring the relative positions, we can go  beyond the weak approximations  used in the past (i.e. $\tilde{\lambda}\ll 1$).
Since we are able to measure the relative positions exactly, we are also able to make these corrections exactly.  
Thus, these relative positions can be used to obtain WVs  even around large values of $\hat{Q}$ that are normally associated with a strong measurement.  
We thus have a much stronger interaction (i.e. a $\lambda $ which does not have to be $\ll 1$) and still we can 
obtain WVs.

\vskip -3cm

\subsubsection{\bf  Obtaining EWVs instead of just an ordinary WV}.
To obtain an EWV (i.e. outside the eigenvalue spectrum), rather than an ordinary WV, we need to control the rotation of the pre- or post-selection
by controlling $\lambda \Delta \hat{Q}$.
E.g. if $Q$ is limited (e.g. $\Delta Q\approx 1$) and $\lambda$ is limited to a particular range sufficient to deliver a EWV  at every point of $Q$ within $\Delta Q$, then we will also obviously obtain a EWV for the quantum average of WVs and do not need to be concerned with other issues such as the slope of $Q$.
However, anytime there is a way to get inside the spectrum of eigenvalues, there will be an exponential increase in the probability to obtain that WV.  
This can be seen from eq. \ref{expweak} in that  the fluctuation in the system is also relevant for the probability to obtain different  post-selections:
as the fluctuation in the system 
increases, the probability of a rare or eccentric post-selection also increases.  However an attempt  to see this through WMs will require 
the spread in the MD to be increased  and this increases the probability of seeing
the strange result as an error of the MD.  

\vskip -2cm

\section{Discussion}

By showing that a single degree of freedom, the total momenta, is shifted by more than it's uncertainty to a WV, we have shown 
that the WV is a definite, non-statistical property that can be associated with every large pre- and post-selected ensemble, not just rare ensembles as was previously believed.  

The new NSWM also presents a limiting case of the relationship between information gain and disturbance of the state. 
Traditionally, it was believed that if a measurement interaction is limited so there is no disturbance on the system, then no information is gained. However, the disturbance on an $N$ particle system with the STWM \S\ref{wvavgop} grows as $N\lambda^2$ while the information gained grows as $N\lambda$.  Considered thus as a limiting process (e.g. if $N\approx \frac{1}{\lambda}$), then the disturbance goes to zero more quickly than the shift in MD.  Thus, if enough particles are used, then information can be obtained even though not even a single particle was disturbed.
The difference between information gained versus disturbance is even more pronounced in the NSWM.

We have thus bolstered the view that the WV is a definite property, rather than a statistical average, because
\begin{itemize}
\item 
no average of eigenvalues can give a EWV outside of the eigenvalue spectrum
\item from a pre-selected-only perspective, it is also not statistical because it is due to a complicated  interference effect involving coordinates of MD.
\end{itemize}

The NSWM can also be used to augment the SWM of \S \ref{wvstat} given a large ensemble ($N\rightarrow\infty$) because we can now interpret what is the average of WVs corresponding to the total momentum shift for a stronger coupling constant by calculating the distribution of pre- and post-selections through the distribution of relative $Q$'s.  We can {\it calculate} what the distribution in $Q$ will be for $N\rightarrow\infty$, even if we do not know the distribution for the individual $Q$'s for any individual particles (the distribution of $\sum Q$ becomes a Gaussian for large $N$ for almost any individual distribution of $Q$). 
However, \cite{abt} for finite $N$ we cannot simply use a calculation because the fluctuation of the relative $Q$'s becomes important and can only be obtained through measurement.

Measurement of the total momentum is an incomplete measurement and thus it's uncertainty will be larger than if it is conditioned on certain deviations of $Q$.  Only for typical distributions of $Q$ can this conditioning be made simply by calculation (rather than measurement).
Thus, measurement of the relative positions provides an advantage when there are significant differences in the distribution for the momentum shift (i.e. finite $N$ and atypical distributions).  

While NSWM is certainly harder to implement experimentally than the traditional WM method described in \S \ref{wvstat}, the instant results are important for conceptual reasons. 
 In addition, it has been shown~\cite{brun} that WMs are more general than projective measurements.  This has resulted in new insights for  many practical applications. 
For example, obtaining the results of \S \ref{wvavgop} without requiring an exponentially rare ensemble and the ability to strengthen the coupling constant could be useful for practical implementations of WMs such as in the amplification of unknown forces~\cite{duck,abt}.  
A full discussion of how the instant NSWM can simultaneously and empirically reveal both the quantum average of WVs and the probability in $Q$ (from eq. \ref{expweak}) is left to a forthcoming publication \cite{abt}.
This gives an empirical grounding for the mathematical techniques utilized in \cite{ab}, thereby extending the work (e.g. ~\cite{at,ab}) and demonstrating the validity of WVs outside of it's original context.

{\bf Acknowledgments:} We thank Alonso Botero for discussions.  YA acknowledges support from the Basic Research Foundation of the Israeli Academy of Sciences.
JT thanks the Templeton foundation for support.

\end{document}